\documentclass[fleqn,usenatbib]{mnras}

\usepackage{newtxtext,newtxmath}
\usepackage[T1]{fontenc}

\DeclareRobustCommand{\VAN}[3]{#2}
\let\VANthebibliography\thebibliography
\def\thebibliography{\DeclareRobustCommand{\VAN}[3]{##3}\VANthebibliography}

\usepackage{graphicx}	
\usepackage{amsmath}	
\usepackage{amssymb}	
\usepackage{gensymb} 

\title[Radio luminosity of GLEAM-X J162759.5-523504.3]{Radio luminosity of GLEAM-X J162759.5-523504.3: does it really exceed the spin-down power of the pulsar?}

\author[M. H. Erkut]{
M. Hakan Erkut$^{1}$\thanks{E-mail: \href{mailto:mherkut@gmail.com}{mherkut@gmail.com}}
\\
$^{1}$Feza Gürsey Center for Physics and Mathematics, Boğaziçi University, 34684, Istanbul, Turkey}

\date{Accepted XXX. Received YYY; in original form ZZZ}

\pubyear{2022}

\begin{document}
\label{firstpage}
\pagerange{\pageref{firstpage}--\pageref{lastpage}}
\maketitle

\begin{abstract}
The recently discovered radio pulsar GLEAM-X J162759.5-523504.3 with an extremely long spin period was reported to have a radio luminosity that exceeds by orders of magnitude the spin-down power of the pulsar. In this Letter, we rigorously calculate the radio luminosity of the source taking into account the dependence of the opening angle of the pulsar-emission cone first on the spin period alone and then on both the spin parameters and the observing frequency. We also revise the value of the spin-down power reported previously. Our analysis is based on the description of the spectral data in terms of two power-law indices as well as a single power-law index. Even if the pulsar's opening angle is treated as a frequency-independent parameter in line with the usual assumption, the period dependence of this parameter implies relatively small opening angles and therefore radio luminosities well below the spin-down power. Although we estimate higher radio luminosities in the physically more plausible case of a frequency-dependent opening angle, the spin-down power is again not exceeded by the highest possible radio luminosity. The radio efficiency of GLEAM-X J162759.5-523504.3 can therefore not be used in favour of a magnetar hypothesis.
\end{abstract}




\section{Introduction} \label{intr}

The rotational energy of the neutron star is generally believed to power the radio emission of pulsars with spin periods ranging from 1.4 ms \citep{2006Sci...311.1901H} to 23.5 s \citep{2018ApJ...866...54T}. Most of the radio pulsars have spin periods $P\lesssim 1$\,s and period derivatives $\dot{P}\lesssim 10^{-13}$\,s\,s$^{-1}$ and span the region above the death line in the $P$--$\dot{P}$ diagram \citep{1975ApJ...196...51R}. The relatively slow radio pulsars of neutron-star spin periods in the $\sim 1$--$20$\,s range consist either of radio emitting transient anomalous X-ray pulsars (AXPs) or of radio pulsars with inferred surface magnetic dipole field strengths being close to or even exceeding the quantum critical value for electrons \citep{2012ApJ...748L..12R}. The long-period radio pulsars such as PSR~J0250+5854 are outnumbered compared to other pulsars partly because of the weak spin-down power which might be inadequate at sufficiently low $\dot{P}$ and high $P$ values to feed the radio emission mechanism in the pulsar's magnetosphere and presumably also because of the selection effects such as low-frequency red noise and short dwell times making the detection of long-period sources difficult in most pulsar surveys \citep{2018ApJ...866...54T}.

The recent discovery of a low-frequency radio transient, namely GLEAM-X J162759.5-523504.3 exhibiting periodic pulsations with an extremely long period of $\sim 1091$\,s has revealed the possible existence of a new class of radio pulsars as an extension of the usual radio pulsar population toward high $P$ and $\dot{P}$ values just above or near the pulsar death line in the $P$--$\dot{P}$ diagram \citep{2022Natur.601..526H}. The radio pulses of GLEAM-X J162759.5-523504.3 are characterized by a stable high linear polarization pointing out the role of well ordered magnetic fields in the origin of radio emission. The typical range for the pseudo-luminosity of the pulses, $L_{\nu}\sim 10^{21}$--$10^{22}$\,erg\,s$^{-1}$\,Hz$^{-1}$, is higher by orders of magnitude than the pseudo-luminosity ranges of flare stars and white-dwarf binaries. The small fractional uncertainty in the measured pulse period strongly indicates the precision of the periodic emission. The search for the period derivative favours $\dot{P}>0$, as expected for most of the radio pulsars \citep{2022Natur.601..526H}. It is therefore highly likely that the radio emitter in GLEAM-X J162759.5-523504.3 is indeed a slowly rotating neutron star and can therefore be seen as a new member of the long-period radio pulsar family.

For most of the radio pulsars, the radio luminosity $L$ is only a small fraction of the pulsar's spin-down power $\dot{E}$ (also known as the spin-down luminosity). An anti-correlation holds between $\dot{E}$ and the so-called radio efficiency $\epsilon \equiv L/\dot{E}$ because of the lack of any notable dependence of $L$ on $\dot{E}$. The vast majority of radio pulsars including binary and millisecond pulsars, radio emitting AXPs, and young pulsars with pulsed high-energy radiation besides normal pulsars have radio efficiencies ranging from $\epsilon \sim 10^{-8}$--$10^{-6}$ when the spin-down power is as high as $\dot{E} \sim 10^{36}$--$10^{37}$\,erg\,s$^{-1}$ to $\epsilon \sim 10^{-3}$--$10^{-1}$ when $\dot{E}<10^{31}$\,erg\,s$^{-1}$ \citep{2014ApJ...784...59S}. For GLEAM-X J162759.5-523504.3, however, $L$ has been claimed to exceed $\dot{E}$ by at least three orders of magnitude in direct contrast with the behaviour of a rotation-powered radio pulsar. Instead, the source has been interpreted as a radio magnetar \citep{2022Natur.601..526H} even though none of the radio emitting magnetar candidates had ever been observed with $\epsilon >1$.

In this Letter, we calculate the radio luminosity of GLEAM-X J162759.5-523504.3 taking into account the dependence of the opening angle of the pulsar beam on the spin period of the neutron star and show that the spin-down power is not exceeded by the radio luminosity of the pulsar. We perform our calculations using the power-law (PL) model first with a single PL index and then with two PL indices to describe the observed average weighted flux density in the $72$--$231$\,MHz frequency range. We also consider the dependence of the opening angle on the observing frequency. In all cases, we find the efficiency of the radio emission from GLEAM-X J162759.5-523504.3 to be consistent with the expected range of $\epsilon$ for pulsars with low $\dot{E}$.

\section{Radio luminosity of the pulsar} \label{lum_pulsar}

\subsection{Frequency-independent opening angle} \label{op_fr_ind}

Assuming that the opening angle of the pulsar beam does not depend on the observing frequency $\nu$, the radio luminosity of a pulsar at a distance $d$ can be written as
\begin{equation}
    L=4\pi d^2 \sin^2\left(\frac{\rho}{2}\right)\int_{\nu_{\rm min}}^{\nu_{\rm max}} S_{\rm p}(\nu)\,{\rm d}\nu,
	\label{lum_def}
\end{equation}
where the so-called opening angle $\rho$ is the angular radius of the pulsar's emission cone and the peak flux density $S_{\rm p}(\nu)$ corresponds to the maximum intensity of the pulse profile \citep{2012hpa..book.....L}. Here, $\sin^2(\rho/2)$ is known as the beaming fraction. The radio pulsars have been observed and probed within the frequency range determined by the integration limits in equation~(\ref{lum_def}). Using the pulse duty cycle $\delta$, the peak flux density $S_{\rm p}$ can be expressed in terms of the mean flux density $S_{\rm m}$. For a flux density spectrum with a PL dependence on the observing frequency $\nu$, we write the mean flux density,
\begin{equation}
    S_{\rm m}(\nu)=S_{\rm m}(\nu_0)\left(\frac{\nu}{\nu_0}\right)^\alpha,
	\label{mean_flux}
\end{equation}
in terms of the reference frequency $\nu_0$ and the PL index $\alpha$. Substituting $S_{\rm p}=S_{\rm m} /\delta$ for the peak flux density and using equation~(\ref{mean_flux}), it follows from equation~(\ref{lum_def}) that
\begin{equation}
    L=4\pi d^2 \sin^2\left(\frac{\rho}{2}\right)\left(\frac{\nu_{\rm max}^{\alpha +1} -\nu_{\rm min}^{\alpha +1}}{\alpha +1}\right) \left[\frac{S_{\rm m}(\nu_0)}{\nu_0^{\alpha}\, \delta}\right].
	\label{lum_1PL}
\end{equation}
Choosing the numerical values, which might be appropriate for most, if not all, radio pulsars, of the parameters in equation~(\ref{lum_1PL}) to be $\delta=0.04$, $\rho=6\degree$ (only for pulsars of spin period $P\sim 1$\,s), $\alpha=-1.8$ \citep[average value of the PL index for pulsar spectra above $100$\,MHz, see][]{2000A&AS..147..195M}, $\nu_{\rm min}=10^7$\,Hz, $\nu_{\rm max}=10^{11}$\,Hz, and $\nu_0=1.4$\,GHz, the radio luminosity of a pulsar can be estimated from equation~(\ref{lum_1PL}) as
\begin{equation}
   L\simeq 7.4\times 10^{30} \left(\frac{d}{{\rm kpc}}\right)^2 \left[\frac{S_{\rm m}(1.4\,{\rm GHz})}{{\rm Jy}}\right]\,{\rm erg\,s^{-1}}
     \label{lum_es}
\end{equation}
\citep{2012hpa..book.....L}. Note that the direct application of equation~(\ref{lum_es}) to the specific case of GLEAM-X J162759.5-523504.3 is not a self-consistent approach to calculate the pulsar's radio luminosity and can be misleading when it comes to radio efficiency. The PL index inferred from the source spectrum is $\alpha=-1.16$ and should be used in equation~(\ref{lum_1PL}) instead of equation~(\ref{lum_es}), which in fact is obtained for $\alpha=-1.8$.

The peak flux density of pulses generated by GLEAM-X J162759.5-523504.3 at a reference frequency of $154$\,MHz was scaled by $\alpha=-1.16$ to the flux density at $1.4$\,GHz in equation~(\ref{lum_es}) to find $L\simeq 4\times 10^{31}$\,erg\,s$^{-1}$ \citep{2022Natur.601..526H}. There are several reasons of why such a high value of $L$, which is well above the spin-down power of the pulsar, was obtained: (i) the flux density at $1.4$\,GHz in equation~(\ref{lum_es}) is the spin-period averaged value of the flux density and cannot be treated as the peak flux density unless the whole expression in equation~(\ref{lum_es}) is multiplied by the pulse duty cycle $\delta$, i.e., $S_{\rm m}=S_{\rm p}\,\delta$, (ii) the flux in equation~(\ref{lum_es}) is overestimated due to the scaling of the flux density by the incompatible PL indices, i.e., the flux is overestimated by $\alpha=-1.8(-1.16)$ at low(high) frequencies, and more importantly (iii) the luminosity estimate by equation~(\ref{lum_es}) is based on the assumption that $P\sim 1$\,s, which leads to the overestimation of the opening angle $\rho$ and thus of $L$ (see equation~\ref{lum_1PL}) for the long-period pulsar GLEAM-X J162759.5-523504.3.

Next, we obtain useful expressions for the radio luminosity of GLEAM-X J162759.5-523504.3 considering first a simple PL model of the source spectrum with a single index $\alpha$ (Section~\ref{single_PL}) and then a joint PL model of the same spectrum with two different PL indices, namely $\alpha$ and $\xi$ (Section~\ref{two_PL}).

\begin{figure}
	\includegraphics[width=\columnwidth]{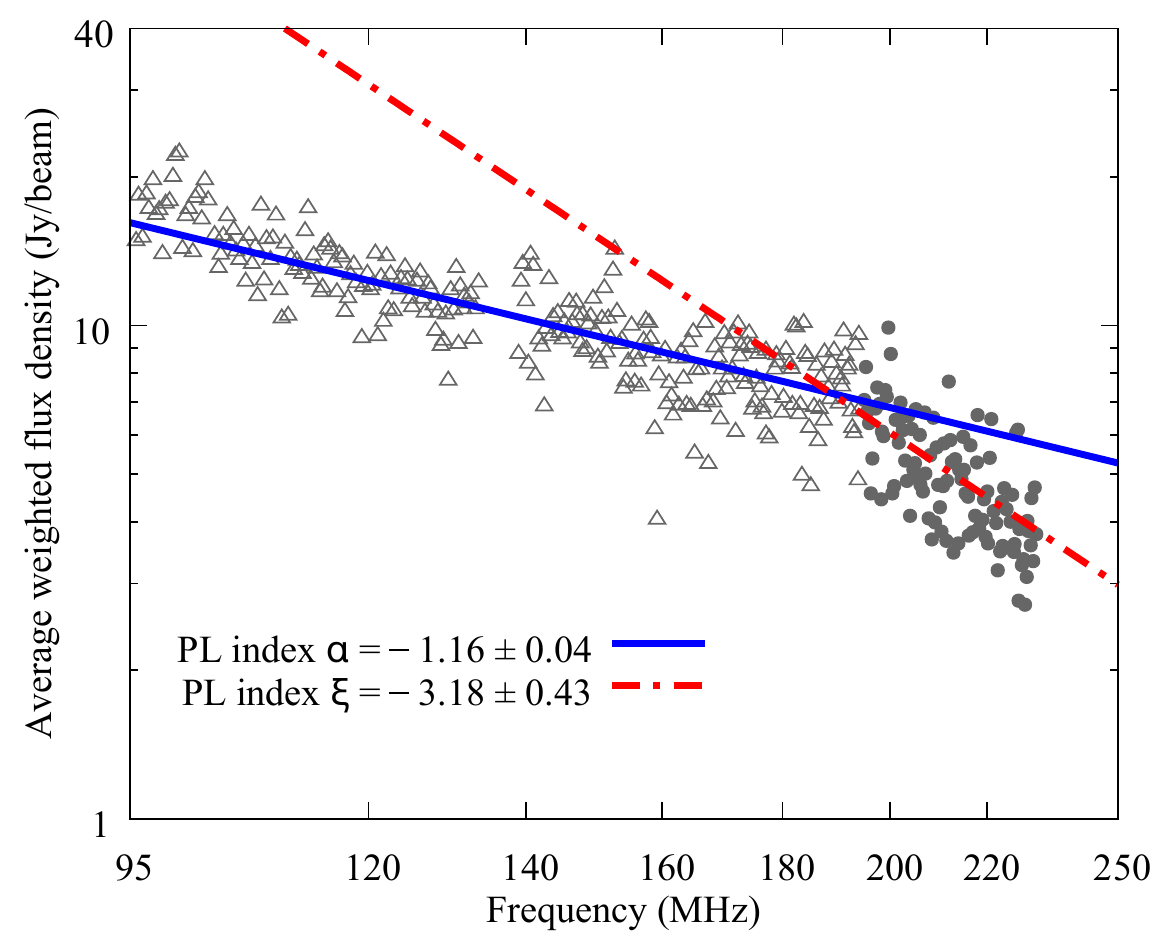}
    \caption{Spectral data fits for GLEAM-X J162759.5-523504.3 using PL models with different indices. Here, we plot the data in \citet{2022Natur.601..526H} without error bars for reasons of clarity. The low-frequency data between $95$ and $195$\,MHz are shown as triangles. Filled circles denote the high-frequency tail of the data above $195$\,MHz. The best fit to the low-frequency part of the data is a PL with an index of $\alpha \simeq -1.16$ (solid blue curve). The best fit to the high-frequency tail of the data between $195$ and $231$\,MHz is another PL with an index of $\xi \simeq -3.18$ (dotted-dashed red curve). See the text for the details about the PL fits in the plot.}
    \label{spectrum}
\end{figure}

\subsubsection{Spectrum with a single PL index}
\label{single_PL}

The PL model fit to the average weighted flux-density data between $95$ and $195$\,MHz yielded a PL spectral index of $\alpha=-1.16\pm 0.04$, which in turn was proposed to represent the extended data between $72$ and $231$\,MHz \citep{2022Natur.601..526H}. The best fit to the spectral data between $95$ and $195$\,MHz is shown in Fig.~\ref{spectrum}.

We calculate the radio luminosity of the pulsar using the expression in equation~(\ref{lum_1PL}) to the extent that the whole spectral data, including the high-frequency tail, can be accounted for using a PL model with a single index $\alpha \simeq -1.16$ (solid blue curve in Fig.~\ref{spectrum}). The pulsar's radio luminosity (see equation~\ref{lum_1PL}) can be written in terms of the peak flux density as
\begin{equation}
    L=4\pi d^2 \sin^2\left(\frac{\rho}{2}\right)\left(\frac{\nu_{\rm max}^{\alpha +1} -\nu_{\rm min}^{\alpha +1}}{\alpha +1}\right) \left[\frac{S_{\rm p}(\nu_0)}{\nu_0^{\alpha}}\right],
	\label{lum_singlePL}
\end{equation}
remembering that $S_{\rm p}=S_{\rm m} /\delta$. For GLEAM-X J162759.5-523504.3, the peak flux density at the reference frequency was reported to be $S_{\rm p}(154\,{\rm MHz})=45$\,Jy \citep[see][]{2022Natur.601..526H}.

\subsubsection{Spectrum with two PL indices}
\label{two_PL}

The average weighted flux-density spectrum was fitted by a power law with an index of $\alpha=-1.16\pm 0.04$ using the least-squares method to find the best fit to the data between $95$ and $195$\,MHz \citep{2022Natur.601..526H}. The PL fit with an index of $\alpha \simeq -1.16$ may, however, overestimate the flux density at the high-frequency tail of the spectrum. The sharp downturn in the flux density at high frequencies ($>195$\,MHz) is noticeable from Fig.~\ref{spectrum}. It is highly likely that a steeper PL fit to the remaining data between $195$ and $231$\,MHz would represent the spectral data extrapolated toward higher frequencies better than a flatter PL with an index of $\sim -1.16$.

We use the nonlinear least-squares Marquardt-Levenberg algorithm to perform a PL fit to the flux-density data in the $195$--$231$\,MHz range. The best fit to the high-frequency data with a reduced $\chi^2$ of $\sim 1.12$ is obtained as a PL with an index of $\xi=-3.18\pm 0.43$ (dotted-dashed red curve in Fig.~\ref{spectrum}). Note that the two PL model fits with indices $\alpha \simeq -1.16$ and $\xi \simeq -3.18$ describe the low and high frequency domains, respectively, and intersect at a frequency $\nu_1 \simeq 189$\,MHz (Fig.~\ref{spectrum}). Next, we consider the frequency of intersection, $\nu_1$, as a limiting point where the spectrum switches the frequency dependence from a relatively flat to a steep PL and therefore as an intermediate limit of the integral in equation~(\ref{lum_def}) to calculate the radio luminosity of the pulsar.

For a spectrum of frequency dependence depicted by two different PL indices, the mean flux density can be expressed as
\begin{equation}
S_{\rm m}(\nu)=
\begin{cases}
  S_{\rm m}(\nu_0)\left(\nu/\nu_0\right)^{\alpha}, \quad & \mbox{for } \nu \leq \nu_1; \\
S_{\rm m}(\nu_1)\left(\nu/\nu_1\right)^{\xi}, \quad & \mbox{for } \nu \geq \nu_1.
\end{cases}
\label{mf_2PL}
\end{equation}
Here, the continuity at $\nu_1$ can only be satisfied if
\begin{equation}
S_{\rm m}(\nu_1)=S_{\rm m}(\nu_0)\left(\frac{\nu_1}{\nu_0}\right)^\alpha.
\label{cnt}
\end{equation}
Using equations~(\ref{mf_2PL}) and (\ref{cnt}) and remembering that $S_{\rm p}=S_{\rm m} /\delta$, it follows from equation~(\ref{lum_def}) that
\begin{equation}
L=L(\nu \leq \nu_1)+L(\nu \geq \nu_1),
\label{lum_2PL}
\end{equation}
where
\begin{equation}
L(\nu \leq \nu_1)=4\pi d^2 \sin^2\left(\frac{\rho}{2}\right)\left(\frac{\nu_1^{\alpha +1} -\nu_{\rm min}^{\alpha +1}}{\alpha +1}\right) \left[\frac{S_{\rm p}(\nu_0)}{\nu_0^{\alpha}}\right]
\label{lum_lsnu1}
\end{equation}
and
\begin{equation}
L(\nu \geq \nu_1)=4\pi d^2 \sin^2\left(\frac{\rho}{2}\right)\left(\frac{\nu_{\rm max}^{\xi +1} -\nu_1^{\xi +1}}{\xi +1}\right) \left[\frac{\nu_1^{\alpha-\xi} S_{\rm p}(\nu_0)}{\nu_0^{\alpha}}\right].
\label{lum_grnu1}
\end{equation}
As expected, equation~(\ref{lum_2PL}) is reduced back to equation~(\ref{lum_singlePL}) when $\xi=\alpha$.

\subsubsection{Spin-period dependence of the opening angle}
\label{result}

The angular radius of the pulsar beam, $\rho$, also known as the opening angle of the pulsar's emission cone, was found to depend on the spin period of the pulsar. The analysis of the emission-cone geometry for a pulsar population revealed the $P^{-1/2}$ dependence of the opening angle for both the so-called inner and outer cones of the radio emission. The relatively long-period pulsars seem to prefer the outer cone with
\begin{equation}
\rho \simeq 6\degree \left(\frac{P}{{\rm s}}\right)^{-1/2},
\label{op_angle}
\end{equation}
where $P$ is the neutron-star spin period \citep{1993ApJ...405..285R,1993A&A...272..268G}. Note that $\rho \simeq 6\degree$ only for pulsars of spin period $\sim 1$\,s. For GLEAM-X J162759.5-523504.3, however, $P\simeq 1091$\,s and therefore $\rho \simeq 0\degree\kern-0.25em.18$ according to equation~(\ref{op_angle}).

\begin{figure}
	\includegraphics[width=\columnwidth]{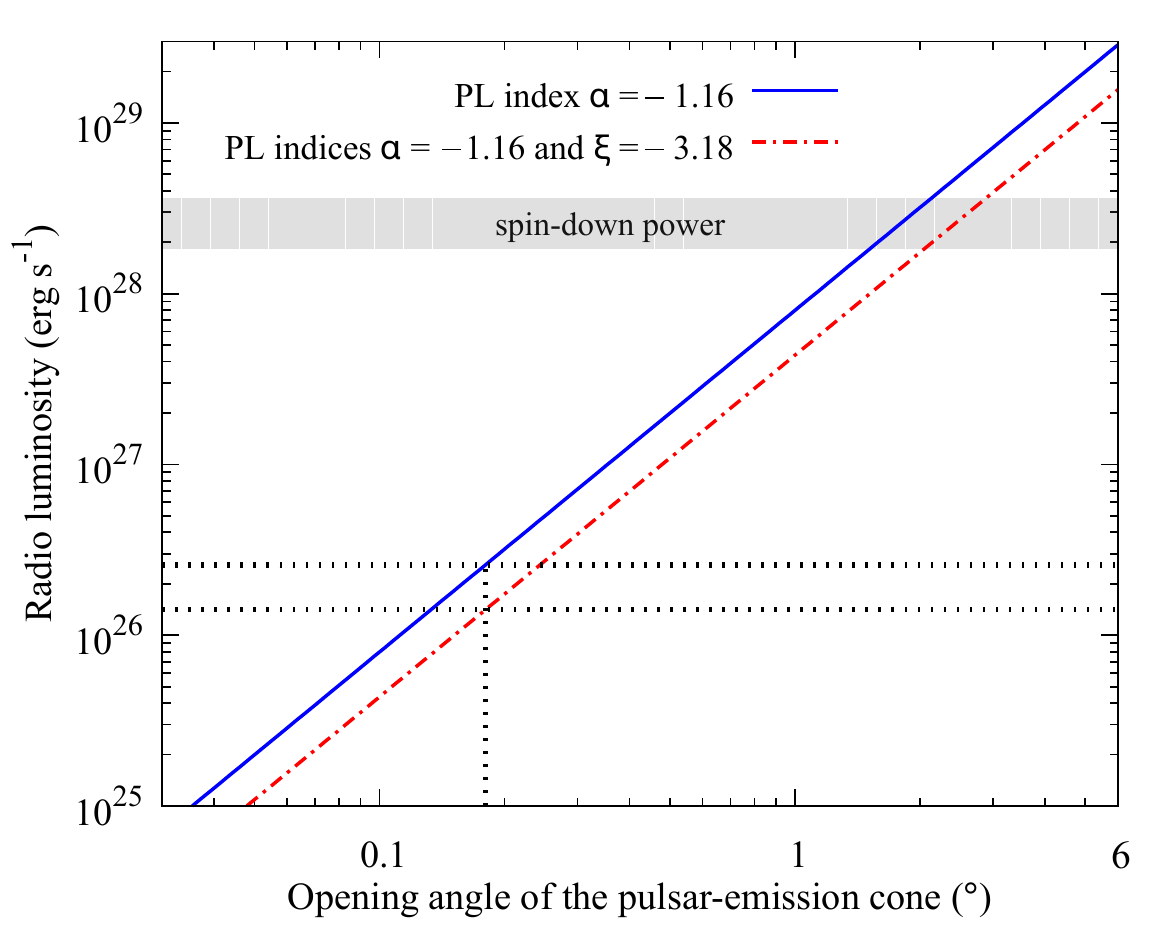}
    \caption{Radio luminosity of GLEAM-X J162759.5-523504.3 as a function of the opening angle $\rho$, which is treated here as independent of $\nu$. The radio luminosity corresponding to the spectrum with a single PL index $\alpha=-1.16$ (equation~\ref{lum_singlePL}) is shown by the solid blue curve. The spectrum described by the two PL indices, $\alpha=-1.16$ and $\xi=-3.18$, yields the radio luminosity (equation~\ref{lum_2PL}) shown by the dotted-dashed red curve. The two possible radio luminosity values of GLEAM-X J162759.5-523504.3 (horizontal dotted lines) are estimated by the solid blue and dotted-dashed red curves at $\rho \simeq 0\degree\kern-0.25em.18$ (see Section~\ref{result}), as shown by the vertical dotted line. The interval for the spin-down power of the pulsar is displayed by the horizontal grey stripe. See the text (Section~\ref{res}) for the specific values of the radio luminosity and the spin-down power.}
    \label{lum}
\end{figure}

For a pulsar's magnetosphere of dipolar field geometry in polar coordinates $(r,\theta)$, $r/\sin^2\theta$ is a constant and can be written for the last open field line at the light-cylinder radius as
\begin{equation}
\frac{r}{\sin^2\theta}=\frac{cP}{2\pi},
\label{dip_geo}
\end{equation}
where $c$ is the speed of light. For regions close to the magnetic axis of the pulsar, that is, for sufficiently small values of $\theta$ and $\rho$ (i.e., $\theta<20\degree$ and $\rho<30\degree$), the relation between the opening angle of the emission cone, $\rho$, and the polar angle of the emission point on the field line, $\theta$, is approximated by $\rho/\theta \simeq 3/2$ \citep{2001ApJ...555...31G}. For small angles, equation~(\ref{dip_geo}) can then be used for finding the relation between $\rho$ and the emission height $r$ as
\begin{equation}
\rho=1\degree\kern-0.25em.24 \left(\frac{r}{10\,{\rm km}}\right)^{1/2} \left(\frac{P}{{\rm s}}\right)^{-1/2}
\label{opang}
\end{equation}
\citep[see also][]{2012hpa..book.....L}. If the emission height $r$ in the pulsar's magnetosphere is independent of the spin period as proposed by \citet{1993ApJ...405..285R}, the empirical relation given by equation~(\ref{op_angle}) can be seen to be in agreement with equation~(\ref{opang}).

The subsequent studies relying on the measurements of pulse width, however, have revealed that the emission height $r$ depends on both the observing frequency $\nu$ and the spin parameters of the pulsar such as the spin period $P$ and its derivative $\dot{P}$ \citep{1997MNRAS.288..631K,2003A&A...397..969K}. The opening angle is then expected in line with equation~(\ref{opang}) to depend on $\nu$ as well. Next, we consider the frequency-dependent opening angle $\rho(\nu)$ and derive, in a similar way to frequency-independent $\rho$ (Sections~\ref{single_PL} and \ref{two_PL}), the luminosity expressions based on the source spectrum to be modelled first with a single PL index and then with two PL indices.

\subsection{Frequency-dependent opening angle} \label{op_fr_dep}

The empirical relationship for the radio emission height in a pulsar magnetosphere can be written in terms of the observing frequency $\nu$ and the spin parameters $P$ and $\dot{P}$ as
\begin{equation}
\frac{r}{10\,{\rm km}}\simeq 40 \left(\frac{\nu}{\nu_{\rm s}}\right)^{\beta} \left(\frac{\dot{P}}{10^{-15}\,{\rm s\,s^{-1}}}\right)^{0.07} \left(\frac{P}{{\rm s}}\right)^{0.30},
\label{em_height}
\end{equation}
where $\beta \simeq -0.26$ and $\nu_{\rm s}=1$\,GHz is a scaling frequency \citep{2003A&A...397..969K}. Note that the radio emission at relatively low frequencies is expected to occur at higher altitudes and therefore with larger opening angles for the pulsar beam (equations~\ref{opang} and \ref{em_height}).

The dependence of the opening angle on the observing frequency can be taken into account when the radio luminosity of a pulsar at a distance $d$,
\begin{equation}
L=4\pi d^2 \int_{\nu_{\rm min}}^{\nu_{\rm max}} \sin^2\left[\frac{\rho(\nu)}{2}\right] S_{\rm p}(\nu)\,{\rm d}\nu,
	\label{lum_gen}
\end{equation}
is calculated through the integration of the peak flux density weighted by the beaming fraction as $\nu$ changes. For sufficiently small values of $\rho$ (see Section~\ref{result}), equation~(\ref{lum_gen}) can be expressed to a very good approximation as
\begin{equation}
L\simeq \pi d^2 \int_{\nu_{\rm min}}^{\nu_{\rm max}} \rho^2(\nu)\, S_{\rm p}(\nu)\,{\rm d}\nu.
	\label{lum_apr}
\end{equation}
Here, $\rho$ is measured in radians.

\subsubsection{Spectrum with a single PL index}
\label{PL_1}

Using equation~(\ref{mean_flux}) with $S_{\rm m}=S_{\rm p}\,\delta$, it follows from equations~(\ref{opang}), (\ref{em_height}), and (\ref{lum_apr}) that
\begin{equation}
    L=4\pi d^2 f(P,\dot{P})\left(\frac{\nu_{\rm max}^{\alpha +\beta +1} -\nu_{\rm min}^{\alpha +\beta +1}}{\alpha +\beta +1}\right) \left[\frac{S_{\rm p}(\nu_0)}{\nu_0^{\alpha}\nu_{\rm s}^{\beta}}\right],
	\label{singlePL_lum}
\end{equation}
where the dimensionless function of $P$ and $\dot{P}$ can be written as
\begin{equation}
f(P,\dot{P})\simeq 4.68\times 10^{-3} \left(\frac{\dot{P}}{10^{-15}\,{\rm s\,s^{-1}}}\right)^{0.07} \left(\frac{P}{{\rm s}}\right)^{-0.7}.
\label{dimf}
\end{equation}
The pulsar-luminosity expression in equation~(\ref{singlePL_lum}) is analogous to that in equation~(\ref{lum_singlePL}) when $\beta$ vanishes.

\subsubsection{Spectrum with two PL indices}
\label{PL_2}

Now, we use equations~(\ref{mf_2PL}) and (\ref{cnt}) with $S_{\rm m}=S_{\rm p}\,\delta$ and obtain, in analogy with equation~(\ref{lum_2PL}), the pulsar luminosity,
\begin{equation}
L=4\pi d^2\left[F(\nu \leq \nu_1)+F(\nu \geq \nu_1)\right],
\label{2PL_lum}
\end{equation}
from equations~(\ref{opang}), (\ref{em_height}), and (\ref{lum_apr}) such that
\begin{equation}
F(\nu \leq \nu_1)=f(P,\dot{P})\left(\frac{\nu_1^{\alpha +\beta +1} -\nu_{\rm min}^{\alpha +\beta +1}}{\alpha +\beta +1}\right) \left[\frac{S_{\rm p}(\nu_0)}{\nu_0^{\alpha}\nu_{\rm s}^{\beta}}\right]
\label{leqnu1}
\end{equation}
and
\begin{equation}
F(\nu \geq \nu_1)=f(P,\dot{P})\left(\frac{\nu_{\rm max}^{\xi +\beta +1} -\nu_1^{\xi +\beta +1}}{\xi +\beta +1}\right) \left[\frac{\nu_1^{\alpha-\xi} S_{\rm p}(\nu_0)}{\nu_0^{\alpha}\nu_{\rm s}^{\beta}}\right].
\label{geqnu1}
\end{equation}
Note that equation~(\ref{2PL_lum}) can be reduced back to equation~(\ref{singlePL_lum}) for $\xi=\alpha$.

Next, we present our results concerning the luminosity estimate of GLEAM-X J162759.5-523504.3 in radio for both cases of frequency-independent and frequency-dependent opening angle.

\begin{figure}
	\includegraphics[width=\columnwidth]{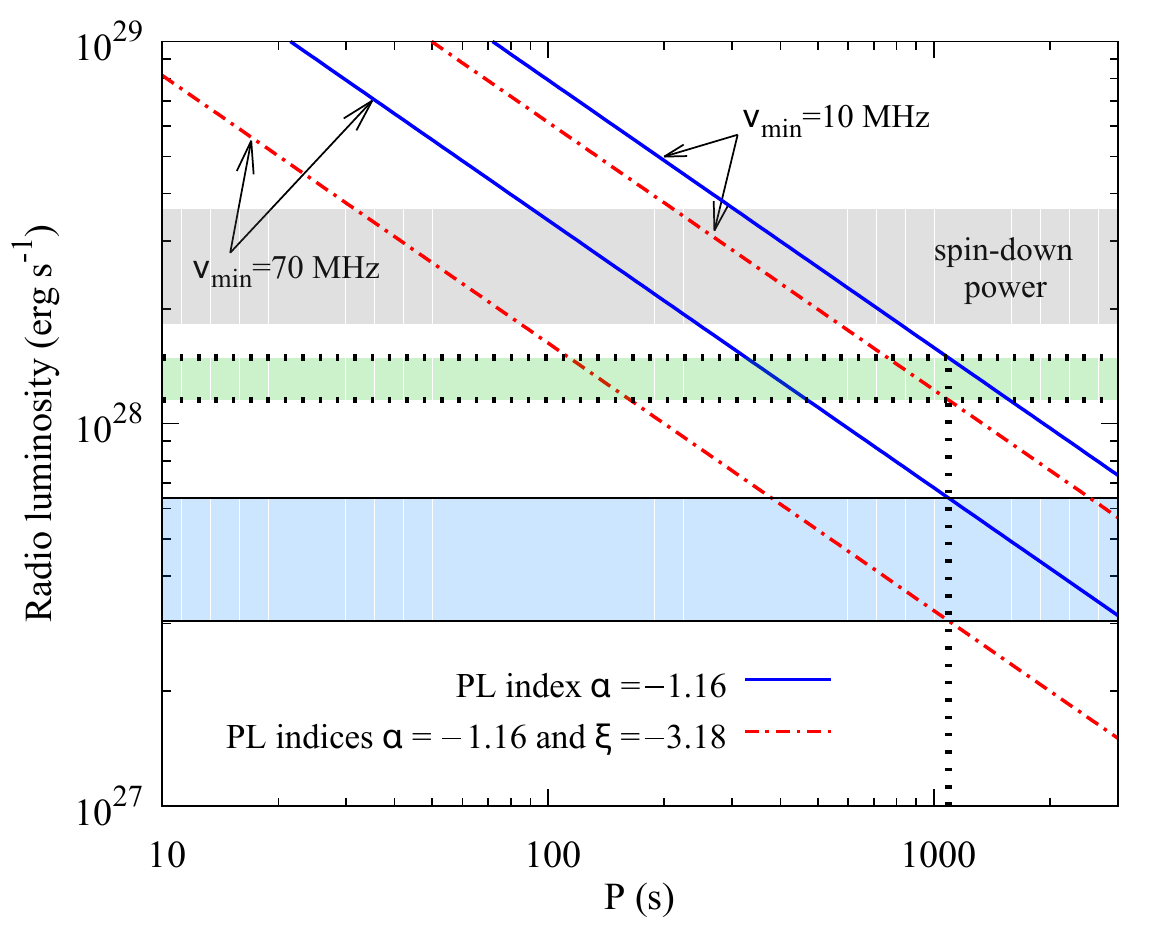}
    \caption{Radio luminosity of GLEAM-X J162759.5-523504.3 as a function of the pulsar's spin period $P$ for an opening angle that depends on $\nu$. The two sets of radio luminosity curves (solid blue and dotted-dashed red curves) are obtained for the two different values of the minimum observing frequency, namely for $\nu_{\rm min}=10$\,MHz and $70$\,MHz. The radio luminosity for a spectrum with a single PL index $\alpha=-1.16$ (equation~\ref{singlePL_lum}) is shown by the solid blue curves. The radio luminosity for a spectrum with two PL indices, $\alpha=-1.16$ and $\xi=-3.18$ (equation~\ref{2PL_lum}), is plotted by the dotted-dashed red curves. The two possible sets of radio luminosity values for GLEAM-X J162759.5-523504.3, which are shown by the horizontal dotted and solid lines as boundaries of green and blue stripes, are estimated at $P\simeq 1091$\,s by the solid blue and dotted-dashed red curves when $\nu_{\rm min}=10$\,MHz and $70$\,MHz, respectively. The interval for the spin-down power of the pulsar is displayed by the horizontal grey stripe as in Fig.~\ref{lum}. See the text (Section~\ref{res}) for the specific values of the radio luminosity and the spin-down power.}
    \label{lumper}
\end{figure}

\section{Results \& discussion} \label{res}

First, we consider the relations derived from the presumption of frequency-independent opening angle (Section~\ref{op_fr_ind}). We employ equation~(\ref{lum_singlePL}) to estimate the radio luminosity of GLEAM-X J162759.5-523504.3 to the extent that the whole spectral data can be described by a single PL index $\alpha \simeq -1.16$ in line with the former suggestion by \citet{2022Natur.601..526H}. We then use equation~(\ref{lum_2PL}) to calculate the radio luminosity of GLEAM-X J162759.5-523504.3 based on a more accurate modelling of the observed spectrum in terms of the two PL indices, $\alpha \simeq -1.16$ and $\xi \simeq -3.18$ to account for the low ($\nu<195$\,MHz) and high ($\nu>195$\,MHz) frequency data, respectively (see Fig.~\ref{spectrum}).

Besides the wide frequency range determined by $\nu_{\rm min}\simeq 10^7$\,Hz and $\nu_{\rm max}\simeq 10^{11}$\,Hz for pulsar study and detection \citep{2012hpa..book.....L}, the numerical values of the parameters we choose in equation~(\ref{lum_singlePL}) and equations~(\ref{lum_2PL})--(\ref{lum_grnu1}) are appropriate to GLEAM-X J162759.5-523504.3. We set $\nu_0=154$\,MHz, $S_{\rm p}(\nu_0)=45$\,Jy, $\alpha=-1.16$, $\xi=-3.18$, and $\nu_1=189$\,MHz.  For a source distance of $d\simeq 1.3$\,kpc \citep{2022Natur.601..526H}, the possible value of the opening angle, i.e., $\rho \simeq 0\degree\kern-0.25em.18$ (see Section~\ref{result}), for $P\simeq 1091$\,s implies a radio luminosity of $L\simeq 2.6\times 10^{26}$\,erg\,s$^{-1}$ for a spectrum with a single PL index and $L\simeq 1.4\times 10^{26}$\,erg\,s$^{-1}$ for a spectrum with two PL indices (Fig.~\ref{lum}).

Even if $\rho$ is treated as a frequency-independent parameter--the direct application of equation~(\ref{lum_es}) by \citet{2022Natur.601..526H} to the radio luminosity of GLEAM-X J162759.5-523504.3 is a typical example--it cannot be regarded as a period-independent parameter. As seen from Fig.~\ref{lum}, the radio luminosity of GLEAM-X J162759.5-523504.3 exceeds the spin-down power of the pulsar,
\begin{equation}
\dot{E}\simeq 1.82\times 10^{28} I_{45}\left(\frac{\dot{P}}{6\times 10^{-10}\,\rm{s\,s^{-1}}}\right)\left(\frac{P}{1091\,{\rm s}}\right)^{-3}\,{\rm erg\,s^{-1}},
\label{sdpow}
\end{equation}
only for $\rho\gtrsim2\degree$, that is, for $P<9$\,s (see equation~\ref{op_angle}). Here, $I_{45}$ is the neutron-star moment of inertia measured in units of $10^{45}$\,g\,cm$^2$. The upper limit for the measured period derivative, $\dot{P}_{{\rm max}}=1.2\times 10^{-9}$\,s\,s$^{-1}$ was found to be twice the best-fit value we use in equation~(\ref{sdpow}) for $\dot{P}$ \citep{2022Natur.601..526H}. The maximum spin-down power of GLEAM-X J162759.5-523504.3 is then $\dot{E}_{{\rm max}}\simeq 3.64\times 10^{28}$\,erg\,s$^{-1}$. The grey stripes plotted in Fig.~\ref{lum} as well as in Fig.~\ref{lumper} are determined by these best-fit and maximum $\dot{E}$ values, which are greater by a factor of $\pi$ than the spin-down luminosity values found by \citet{2022Natur.601..526H} who probably missed such a factor while computing the spin-down power. As a long-period pulsar, GLEAM-X J162759.5-523504.3 is expected to have a fainter radio luminosity with respect to the spin-down power than an ordinary pulsar of $P\sim 1$\,s with $\rho \sim 6\degree$ (Fig.~\ref{lum}).

In a physically more plausible scenario, however, $\rho$ increases with decreasing $\nu$ as suggested by equations~(\ref{opang}) and (\ref{em_height}) and we need to consider the relations based on the frequency-dependent opening angle (Section~\ref{op_fr_dep}). We basically use equations~(\ref{singlePL_lum}) and (\ref{2PL_lum}) to estimate the radio luminosity of GLEAM-X J162759.5-523504.3 for a spectrum with a single PL index and a spectrum with two PL indices, respectively.

In addition to all numerical values we have chosen in the case of frequency-independent $\rho$ for the parameters such as $\nu_0$, $S_{\rm p}(\nu_0)$, $\alpha$, $\xi$, $\nu_1$, and $d$, we set $\nu_{\rm s}=1$\,GHz, $\beta=-0.26$, and $\dot{P}=6\times 10^{-10}$\,s\,s$^{-1}$ to obtain the luminosity estimate of GLEAM-X J162759.5-523504.3, as shown in Fig.~\ref{lumper}. The luminosity curves are plotted for two different values of $\nu_{{\rm min}}$ to see the contribution of ever increasing $\rho$ at low frequencies to the total radio output of the source. In the absence of any spectral data of GLEAM-X J162759.5-523504.3 at frequencies below 70 MHz, we consider two possibilities: (i) the radio emission might be extended to very low frequencies, i.e., $\nu_{{\rm min}}\simeq 10$\,MHz, but might have suffered from interstellar and ionospheric scintillation or (ii) the source is not radio luminous at all at frequencies less than $\nu_{{\rm min}}\simeq 70$\,MHz. In Fig.~\ref{lumper}, the curves labelled with $\nu_{{\rm min}}=10$\,MHz yield the highest possible radio luminosity values at $P\simeq 1091$\,s and yet the spin-down power is not exceeded. We read $L\simeq 1.5\times 10^{28}$\,erg\,s$^{-1}$ for a single PL index and $L\simeq 1.2\times 10^{28}$\,erg\,s$^{-1}$ for double PL indices. The radio luminosities estimated by the curves labelled with $\nu_{{\rm min}}=70$\,MHz are $L\simeq 6.4\times 10^{27}$\,erg\,s$^{-1}$ and $L\simeq 3.0\times 10^{27}$\,erg\,s$^{-1}$ for a single PL index and double PL indices, respectively (Fig.~\ref{lumper}).

For a radio pulsar of $P$ exceeding $10^3$\,s such as GLEAM-X J162759.5-523504.3, the opening angles are expected to attain values as small as $\rho \simeq 0\degree\kern-0.25em.18$ for the frequency-independent case. Albeit higher, the opening angle in equation~(\ref{lum_apr}) for the frequency-dependent case can be seen from equations~(\ref{opang}) and (\ref{em_height}) to vary between $\rho_{{\rm max}}\simeq 2\degree$ and $\rho_{{\rm min}}\simeq 0\degree\kern-0.25em.6$ at $\nu_{{\rm min}}=10^7$\,Hz and $\nu_{{\rm max}}=10^{11}$\,Hz, respectively. There is, however, a lower limit for $\rho$ due to relativistic beaming and the realization of such small $\rho$ values must be checked. Indeed, $\rho>\Delta \rho$, where $\Delta \rho \simeq \gamma^{-1}$ is the opening angle of the curvature-radiation cone generated by a localized source of ultra-relativistic particles of Lorentz factor $\gamma$ moving along dipolar magnetic field lines. The typical value of $\gamma$ in a pulsar magnetosphere is of the order of 100, i.e., $10^2<\gamma<10^3$ \citep{1983A&A...123....7G,2004ApJ...609..335G,2005A&A...432L..61G}, yielding the $0\degree\kern-0.25em.57$--$0\degree\kern-0.25em.057$ range for $\Delta \rho$. Our range for the frequency-dependent $\rho$ values is therefore already plausible. The smallest value of the lower limit for $\gamma$ can be estimated using equation~(6) in \citet{1983A&A...123....7G}. Substituting $220$\,MHz for the maximum emitted frequency, 10 or 100 for the maximum coherence parameter $\kappa$, and $1227$ for the $f=r_{{\rm min}} /10\,{\rm km}$ parameter corresponding to $\nu=220$\,MHz (see equation~\ref{em_height}) together with other parameters being appropriate to GLEAM-X J162759.5-523504.3, we obtain the lower limits for $\gamma$ as $\gamma_{{\rm min}}\simeq 696$ when $\kappa=10$ and $\gamma_{{\rm min}}\simeq 323$ when $\kappa=100$. The corresponding $\Delta \rho$ values are $< 0\degree\kern-0.25em.08$ and $0\degree\kern-0.25em.18$, respectively. Even if $\rho$ slightly exceeds $0\degree\kern-0.25em.18$ for smaller values of $\gamma$, the radio emission can still be rotationally powered since $L<\dot{E}$ for $\rho \lesssim 2\degree$ (see Fig.~\ref{lum}).

In summary, the highest possible radio luminosity of GLEAM-X J162759.5-523504.3 based on the maximum peak flux density of $\sim 45$\,Jy observed at $\nu_0=154$\,MHz does not exceed the pulsar's spin-down power of $\sim 1.8\times 10^{28}$\,erg\,s$^{-1}$. It is rather highly likely that the source is a pulsating neutron star with a radio efficiency $\epsilon \gtrsim 0.1$ as expected from the long-period members of the pulsar population with sufficiently low spin-down power (see Section~\ref{intr}). The opening angles of such long-period pulsars are extremely small at relatively high frequencies making the detection of such transient sources more likely at low frequencies.


\section*{Acknowledgements}

I thank O. Çatmabacak and Ö. Çatmabacak for inspiring discussions and useful comments.

\section*{Data Availability}

The data used in this article are already available in \citet{2022Natur.601..526H}.



\bibliographystyle{mnras}
\bibliography{ref} 

\begin{thebibliography}{}
\makeatletter
\relax
\def\mn@urlcharsother{\let\do\@makeother \do\$\do\&\do\#\do\^\do\_\do\%\do\~}
\def\mn@doi{\begingroup\mn@urlcharsother \@ifnextchar [ {\mn@doi@}
  {\mn@doi@[]}}
\def\mn@doi@[#1]#2{\def\@tempa{#1}\ifx\@tempa\@empty \href
  {http://dx.doi.org/#2} {doi:#2}\else \href {http://dx.doi.org/#2} {#1}\fi
  \endgroup}
\def\mn@eprint#1#2{\mn@eprint@#1:#2::\@nil}
\def\mn@eprint@arXiv#1{\href {http://arxiv.org/abs/#1} {{\tt arXiv:#1}}}
\def\mn@eprint@dblp#1{\href {http://dblp.uni-trier.de/rec/bibtex/#1.xml}
  {dblp:#1}}
\def\mn@eprint@#1:#2:#3:#4\@nil{\def\@tempa {#1}\def\@tempb {#2}\def\@tempc
  {#3}\ifx \@tempc \@empty \let \@tempc \@tempb \let \@tempb \@tempa \fi \ifx
  \@tempb \@empty \def\@tempb {arXiv}\fi \@ifundefined
  {mn@eprint@\@tempb}{\@tempb:\@tempc}{\expandafter \expandafter \csname
  mn@eprint@\@tempb\endcsname \expandafter{\@tempc}}}

\bibitem[\protect\citeauthoryear{{Gangadhara}}{{Gangadhara}}{2004}]{2004ApJ...609..335G}
{Gangadhara} R.~T.,  2004, \mn@doi [\apj] {10.1086/420961}, \href
  {https://ui.adsabs.harvard.edu/abs/2004ApJ...609..335G} {609, 335}

\bibitem[\protect\citeauthoryear{{Gangadhara} \& {Gupta}}{{Gangadhara} \&
  {Gupta}}{2001}]{2001ApJ...555...31G}
{Gangadhara} R.~T.,  {Gupta} Y.,  2001, \mn@doi [\apj] {10.1086/321439}, \href
  {https://ui.adsabs.harvard.edu/abs/2001ApJ...555...31G} {555, 31}

\bibitem[\protect\citeauthoryear{{Gil}}{{Gil}}{1983}]{1983A&A...123....7G}
{Gil} J.,  1983, \aap, \href
  {https://ui.adsabs.harvard.edu/abs/1983A&A...123....7G} {123, 7}

\bibitem[\protect\citeauthoryear{{Gil} \& {Melikidze}}{{Gil} \&
  {Melikidze}}{2005}]{2005A&A...432L..61G}
{Gil} J.,  {Melikidze} G.~I.,  2005, \mn@doi [\aap]
  {10.1051/0004-6361:200500021}, \href
  {https://ui.adsabs.harvard.edu/abs/2005A&A...432L..61G} {432, L61}

\bibitem[\protect\citeauthoryear{{Gil}, {Kijak}  \& {Seiradakis}}{{Gil}
  et~al.}{1993}]{1993A&A...272..268G}
{Gil} J.~A.,  {Kijak} J.,   {Seiradakis} J.~H.,  1993, \aap, \href
  {https://ui.adsabs.harvard.edu/abs/1993A&A...272..268G} {272, 268}

\bibitem[\protect\citeauthoryear{{Hessels}, {Ransom}, {Stairs}, {Freire},
  {Kaspi}  \& {Camilo}}{{Hessels} et~al.}{2006}]{2006Sci...311.1901H}
{Hessels} J. W.~T.,  {Ransom} S.~M.,  {Stairs} I.~H.,  {Freire} P. C.~C.,
  {Kaspi} V.~M.,   {Camilo} F.,  2006, \mn@doi [Science]
  {10.1126/science.1123430}, \href
  {https://ui.adsabs.harvard.edu/abs/2006Sci...311.1901H} {311, 1901}

\bibitem[\protect\citeauthoryear{{Hurley-Walker} et~al.,}{{Hurley-Walker}
  et~al.}{2022}]{2022Natur.601..526H}
{Hurley-Walker} N.,  et~al., 2022, \mn@doi [\nat] {10.1038/s41586-021-04272-x},
  \href {https://ui.adsabs.harvard.edu/abs/2022Natur.601..526H} {601, 526}

\bibitem[\protect\citeauthoryear{{Kijak} \& {Gil}}{{Kijak} \&
  {Gil}}{1997}]{1997MNRAS.288..631K}
{Kijak} J.,  {Gil} J.,  1997, \mn@doi [\mnras] {10.1093/mnras/288.3.631}, \href
  {https://ui.adsabs.harvard.edu/abs/1997MNRAS.288..631K} {288, 631}

\bibitem[\protect\citeauthoryear{{Kijak} \& {Gil}}{{Kijak} \&
  {Gil}}{2003}]{2003A&A...397..969K}
{Kijak} J.,  {Gil} J.,  2003, \mn@doi [\aap] {10.1051/0004-6361:20021583},
  \href {https://ui.adsabs.harvard.edu/abs/2003A&A...397..969K} {397, 969}

\bibitem[\protect\citeauthoryear{{Lorimer} \& {Kramer}}{{Lorimer} \&
  {Kramer}}{2012}]{2012hpa..book.....L}
{Lorimer} D.~R.,  {Kramer} M.,  2012, {Handbook of Pulsar Astronomy}.
Cambridge University Press

\bibitem[\protect\citeauthoryear{{Maron}, {Kijak}, {Kramer}  \&
  {Wielebinski}}{{Maron} et~al.}{2000}]{2000A&AS..147..195M}
{Maron} O.,  {Kijak} J.,  {Kramer} M.,   {Wielebinski} R.,  2000, \mn@doi
  [\aaps] {10.1051/aas:2000298}, \href
  {https://ui.adsabs.harvard.edu/abs/2000A&AS..147..195M} {147, 195}

\bibitem[\protect\citeauthoryear{{Rankin}}{{Rankin}}{1993}]{1993ApJ...405..285R}
{Rankin} J.~M.,  1993, \mn@doi [\apj] {10.1086/172361}, \href
  {https://ui.adsabs.harvard.edu/abs/1993ApJ...405..285R} {405, 285}

\bibitem[\protect\citeauthoryear{{Rea}, {Pons}, {Torres}  \& {Turolla}}{{Rea}
  et~al.}{2012}]{2012ApJ...748L..12R}
{Rea} N.,  {Pons} J.~A.,  {Torres} D.~F.,   {Turolla} R.,  2012, \mn@doi
  [\apjl] {10.1088/2041-8205/748/1/L12}, \href
  {https://ui.adsabs.harvard.edu/abs/2012ApJ...748L..12R} {748, L12}

\bibitem[\protect\citeauthoryear{{Ruderman} \& {Sutherland}}{{Ruderman} \&
  {Sutherland}}{1975}]{1975ApJ...196...51R}
{Ruderman} M.~A.,  {Sutherland} P.~G.,  1975, \mn@doi [\apj] {10.1086/153393},
  \href {https://ui.adsabs.harvard.edu/abs/1975ApJ...196...51R} {196, 51}

\bibitem[\protect\citeauthoryear{{Szary}, {Zhang}, {Melikidze}, {Gil}  \&
  {Xu}}{{Szary} et~al.}{2014}]{2014ApJ...784...59S}
{Szary} A.,  {Zhang} B.,  {Melikidze} G.~I.,  {Gil} J.,   {Xu} R.-X.,  2014,
  \mn@doi [\apj] {10.1088/0004-637X/784/1/59}, \href
  {https://ui.adsabs.harvard.edu/abs/2014ApJ...784...59S} {784, 59}

\bibitem[\protect\citeauthoryear{{Tan} et~al.,}{{Tan}
  et~al.}{2018}]{2018ApJ...866...54T}
{Tan} C.~M.,  et~al., 2018, \mn@doi [\apj] {10.3847/1538-4357/aade88}, \href
  {https://ui.adsabs.harvard.edu/abs/2018ApJ...866...54T} {866, 54}

\makeatother
\end{thebibliography}








\bsp	
\label{lastpage}
\end{document}